\documentclass[%
 reprint,
 amsmath,amssymb,
 aps,
]{revtex4-2}

\usepackage{graphicx,amsmath,amssymb,amsfonts,latexsym,color,dcolumn,bm,mathtools,dsfont}
\usepackage{verbatim,epstopdf}
\usepackage[colorlinks=true,hidelinks]{hyperref}
\usepackage{siunitx}

\providecommand{\moy}[1]{\langle #1 \rangle}

\providecommand{\ket}[1]{\lvert #1 \rangle}

\definecolor{red}{rgb}{0,0,0}
\newcommand{\red}[1]{\textcolor{red}{#1}}

\definecolor{blue}{rgb}{0,0,0}
\newcommand{\blue}[1]{\textcolor{blue}{#1}}

\begin{document}

\title{Efficient qubit measurement with a nonreciprocal microwave amplifier}

\author{F. Lecocq$^{1,2}$, L. Ranzani$^3$, G. A. Peterson$^{1,2}$, K. Cicak$^1$, X. Y. Jin$^{1,2}$, R. W. Simmonds$^1$, J. D. Teufel$^1$ and J. Aumentado$^1$}
\email{florent.lecocq@nist.gov}
\email{jose.aumentado@nist.gov}
\affiliation{$^1$National Institute of Standards and Technology, 325 Broadway, Boulder, CO 80305, USA}
\affiliation{$^2$Dept. of Physics, University of Colorado, 2000 Colorado Ave., Boulder, CO 80309, USA}
\affiliation{$^3$Raytheon BBN Technologies, Cambridge, MA 02138, USA}

\date{\today}% It is always \today, today,

%\begin{abstract}
%The measurement of a superconducting quantum bit is often performed by encoding its state in a single quadrature of a microwave field. Ideal measurement efficiency of this observable could in principle be achieved by noiselessly amplifying the information-carrying quadrature, but in practice is limited by technical losses due to circulators, cables and connectors used in state-of-the-art amplification chains. In this work we approach ideal measurement efficiency by directly connecting a 3D transmon to a non-reciprocal phase-sensitive amplifier. We achieve a measurement efficiency in excess of $75~\%$ while maintaining a spurious cavity population due to amplifier backaction below $0.1~\text{photons}$.
%\end{abstract}

\begin{abstract}
The act of observing a quantum object fundamentally perturbs its state, resulting in a random walk toward an eigenstate of the measurement operator. Ideally, the measurement is responsible for all dephasing of the quantum state. In practice, imperfections in the measurement apparatus limit or corrupt the flow of information required for quantum feedback protocols, an effect quantified by the measurement efficiency. Here we demonstrate the efficient measurement of a superconducting qubit using a nonreciprocal parametric amplifier to directly monitor the microwave field of a readout cavity. By mitigating the losses between the cavity and the amplifier we achieve a measurement efficiency of $72\%$. The directionality of the amplifier protects the readout cavity and qubit from excess backaction caused by amplified vacuum fluctuations. In addition to providing tools for further improving the fidelity of strong projective measurement, this work creates a testbed for the experimental study of ideal weak measurements, and it opens the way towards quantum feedback protocols based on weak measurement such as state stabilization or error correction.
\end{abstract}

\maketitle

Quantum measurements often involve entangling the system of interest with light \cite{Braginsky1992QuantumMeasurement}. As a consequence, the subsequent measurement of the light, performed by an observer or by the unmonitored environment, affects the quantum state of the system \cite{Haroche2006}. The measurement efficiency, $0 \leq \eta \leq 1$, characterizes the fraction of the total available quantum state information that is acquired by the observer. Using a quantum non-demolition measurement scheme, highly accurate state estimation can be performed by repeating many inefficient measurements \cite{Harty2014High-FidelityBit,Elder2020High-FidelityCircuits,Todaro2020StateDetector}. A higher efficiency results in measurement speed up and therefore contributes to the fidelity of the state estimation, a crucial metric for quantum computation \cite{Walter2017RapidQubits,Arute2019QuantumProcessor}. Meanwhile, an efficient measurement is critical for measurement-based quantum control \cite{Wiseman2009QuantumControl}, in which the measurement outcome is used to feed back on the quantum system, enabling, for example, quantum state stabilization \cite{Sayrin2011Real-timeStates,Vijay2012,deLange2014ReversingFeedback,Rossi2018Measurement-basedMotion} or measurement-based entanglement \cite{Roch2014}.

A perfectly efficient measurement requires both ideal collection of the light and a faithful, noiseless detector. For superconducting qubits, the use of a microwave readout resonator with a one-dimensional radiation pattern enables the collection of every photon \cite{Houck2007}. A noiseless detector must only measure the information-carrying quadrature of the microwave field \red{and,} in principle, this can be implemented with a phase-sensitive parametric amplifier \cite{Caves1982QuantumAmplifiers}. In reality, this approach requires additional hardware to operate which introduces unavoidable losses in the signal path, reducing the overall efficiency. In particular, these amplifiers are reciprocal and therefore rely on magnetic microwave circulators to control the signal flow, leading to significant component and wiring losses \cite{Ranzani2013,Aumentado2020SuperconductingAmplifiers}. In recent years, various nonreciprocal alternatives to conventional parametric amplifiers have been developed, breaking reciprocity via parametric interactions \cite{Ranzani2015,Metelmann2015,Sliwa2015,Chapman2017WidelyCircuits,Lecocq2017NonreciprocalAmplifier,Lecocq2020MicrowaveAmplifier}, time domain operations \cite{Rosenthal2020EfficientDetector}, or traveling wave amplification \cite{Macklin2015AAmplifier}. \red{However,} the application of these nonreciprocal amplifiers to efficient superconducting qubit measurement is still a nascent field \cite{Abdo2014,Thorbeck2017ReverseAmplifier,Eddins2019High-EfficiencyAmplifier,Rosenthal2020EfficientDetector,Abdo2020On-chipMeasurements} with many outstanding experimental and theoretical challenges. \blue{Indeed, in the absence of an intermediate circulator, the amplifier and the device under test can no longer be treated as independent modular elements, in stark contrast with conventional parametric amplification chains. In fact, the devices merge into a new and unique quantum system, which introduces the difficulty of combining large amplified signals with fragile quantum states. In this work we solve this delicate balance by using coupled mode theory to finely tune \textit{in situ} parametric interference within this larger quantum system.}

\begin{figure}[h]
	\includegraphics[scale=1.0]{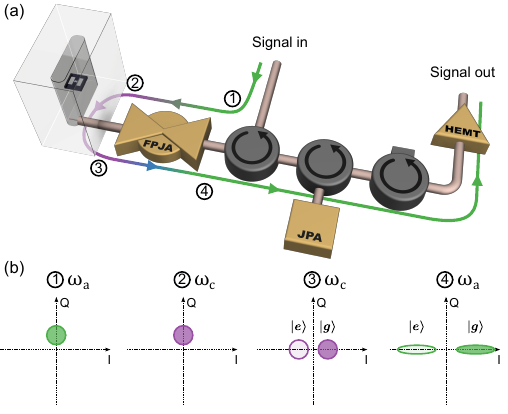}
	\caption{(a) Diagram of the high efficiency qubit measurement chain. A transmon qubit is dispersively coupled to a 3D aluminum readout cavity. The readout cavity is directly connected  to a nonreciprocal phase-sensitive amplifier based on an FPJA. (b) Phase-space representation of the microwave field at different part of the circuit. An input measurement drive is upconverted to the cavity frequency then reflected with a qubit state dependent phase shift. The quadrature containing the qubit state information is amplified and the signal is downconverted and routed out for further amplification. Both the FPJA and the JPA have an on-chip dc and ac flux bias (not shown). The qubit is driven via a weakly coupled cavity port (not shown).
	\label{fig1}}
\end{figure}
 Here we \red{couple} the readout cavity of a transmon qubit to a Field Programmable Josephson Amplifier (FPJA) \cite{Lecocq2017NonreciprocalAmplifier}, see Fig.~\ref{fig1}. We operate the FPJA as a directional phase-sensitive parametric amplifier \cite{Lecocq2020MicrowaveAmplifier} to monitor only the quadrature of the microwave field in which the qubit state is encoded. The directionality of the amplifier protects the cavity from amplified vacuum fluctuations, circumventing the need for conventional microwave circulators between the cavity and the amplifier. As a consequence of the low loss and low added noise in this system, we achieve a measurement efficiency of $72\%$, \red{amongst the best reported in the literature \cite{Walter2017RapidQubits,Eddins2019High-EfficiencyAmplifier,Rosenthal2020EfficientDetector}.}

A transmon qubit, with a resonance frequency $\omega_\text{q}/2\pi\approx6.297~\si{\giga\hertz}$ \red{and relaxation time $T_1=27~\si{\micro \second}$}, is embedded inside a three-dimensional aluminum readout cavity \cite{Paik2011} with a resonant frequency, $\omega_\text{r}/2\pi\approx10.929~\si{\giga\hertz}$. The capacitive coupling between the qubit and the cavity results in a qubit state dependent cavity resonance frequency: $\omega_\text{r}+\chi$ or $\omega_\text{r}-\chi$ for the qubit respectively in the ground state $\ket{g}$ or excited state $\ket{e}$, where $2\chi/2\pi\approx1.7~\si{\mega\hertz}$. The FPJA is part of a class of multi-mode amplifiers whose behavior can be programmed \textit{in-situ} by a set of parametric drives \cite{Ranzani2015,Metelmann2015,Sliwa2015,Lecocq2017NonreciprocalAmplifier}. In the following, we use four parametric drives to program the FPJA as a directional phase-sensitive amplifier  \cite{Lecocq2020MicrowaveAmplifier}. The FPJA has three flux-tunable resonances: the input resonator is tuned into resonance with the readout cavity, $\omega_\text{c}=\omega_\text{r}$, resulting in an output resonator frequency $\omega_\text{a}/2\pi\approx6.912~\si{\giga\hertz}$ and an internal amplification resonator frequency $\omega_\text{b}/2\pi\approx8.013~\si{\giga\hertz}$ (Supp. Inf.). The strength and phase of the each pump is set to create the following nonreciprocal signal flow: (1) A signal enters the FPJA output port, at a frequency $\omega_\text{a}$, (2) the signal is upconverted to the input port of the FPJA, at a frequency $\omega_\text{c}$, (3) the signal reflects off the readout cavity with a qubit state dependent phase shift and re-enters the FPJA input port, (4) the signal is downconverted to the amplification resonator, at a frequency $\omega_\text{b}$, and the quadrature containing the qubit state information is amplified with a gain $G$ tunable \textit{in situ} before further downconversion back to the output port, at a frequency $\omega_\text{a}$. Finally, the signal is routed to a conventional homodyne measurement setup. To characterize this system's measurement efficiency we use a robust method that relies on the comparison of the qubit dephasing rate to the measurement rate \cite{Bultink2018GeneralQED, Eddins2019High-EfficiencyAmplifier, Touzard2019GatedQubits, Rosenthal2020EfficientDetector}.

\begin{figure}
	\includegraphics[scale=1.0]{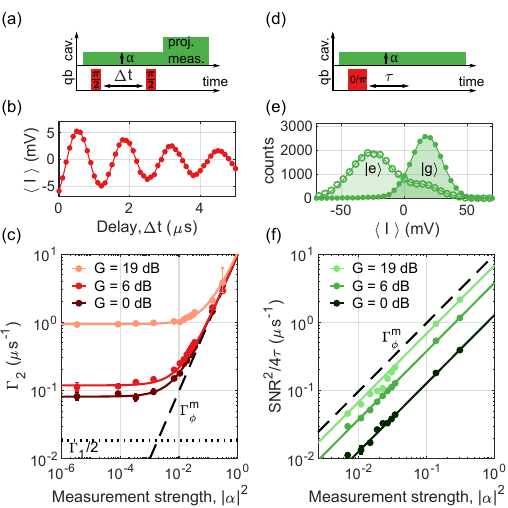}
	\caption{Measurement induced dephasing and measurement rate.  (a) Pulse sequence to measure Ramsey oscillations in presence of a measurement of variable strength $|\alpha|^2$. (b) Example of a Ramsey oscillation with an exponential decay rate $\Gamma_2$, for $G\approx6~\text{dB}$ and $|\alpha|^2\approx0.01$. (c) Qubit decoherence rate $\Gamma_2$ as a function of measurement strength $|\alpha|^2$ and amplifier gain $G$. (d) Pulse sequence to measure the signal to noise ratio between ground and excited qubit state as a function of the measurement strength $|\alpha|^2$. (e) Typical histogram of the measurement signal for the qubit prepared in the ground or excited state, for $G\approx15~\text{dB}$ and $|\alpha|^2\approx0.3$. (f) Measurement rate, $\Gamma_m = \text{SNR}^2/4\tau$, as a function of measurement strength $|\alpha|^2$ and amplifier gain $G$. In (b), (c), (e) and (f) the dots are data and the solid lines are linear fits.
	\label{fig2}}
\end{figure}

We first calibrate the qubit measurement strength by observing the associated measurement-induced dephasing, see Fig.~\ref{fig2} a-c. The total qubit decoherence rate is $\Gamma_2=\Gamma_1/2+\Gamma_\phi^\text{env}+\Gamma_\phi^\text{m}$ where $\Gamma_1$ is the relaxation rate, $\Gamma_\phi^\text{m}$ is the measurement-induced dephasing rate and $\Gamma_\phi^\text{env}$ is the excess dephasing rate that is not due to the measurement. Both dephasing rates can be expressed as a function of the number of photons in the cavity \cite{Gambetta2008QuantumEffect,Wang2019CavityQubits}:

\begin{align}
\begin{split}
\Gamma_\phi^\text{env}=\frac{4\chi^2\kappa}{4\chi^2+\kappa^2}n_\text{env},\\
\Gamma_\phi^\text{m}=\frac{8\chi^2\kappa}{4\chi^2+\kappa^2}|\alpha|^2,
\end{split}
\label{EqGammaPhi}
\end{align}

where $\kappa/2\pi\approx2.58~\si{\mega\hertz}$ is the effective cavity linewidth, $n_\text{env}$ is the effective thermal occupancy of the cavity and $|\alpha|^2$ is the mean number of coherent photons that parameterizes the measurement strength. Using the Ramsey sequence shown in Fig.~\ref{fig2}a we measure the qubit decoherence rate $\Gamma_2$ as a function of the measurement strength $|\alpha|^2$ and of the FPJA gain $G$, shown in Fig.~\ref{fig2}c. \red{The measurement strength $|\alpha|^2$ is calibrated using Eq.~\ref{EqGammaPhi} and separate measurements of $\kappa$ and $\chi$}.  At low measurement strength, the qubit decoherence rate is mostly dominated by excess dephasing $\Gamma_\phi^\text{env}$. At zero gain, $\Gamma_\phi^\text{env}$ originates from residual thermal photons in the cavity \cite{Wang2019CavityQubits}. At higher gain, due to the finite directionality of the FPJA, residual amplified vacuum fluctuations drive the readout cavity and slightly increase the excess dephasing. The decoherence rate then increases with measurement strength, and even at the highest gain we maintain a measurement-induced dephasing rate that dominates the decoherence rate, $\Gamma_\phi^\text{m}\gg(\Gamma_1/2+\Gamma_\phi^\text{env})$.

We now extract the signal to noise ratio of the measurement chain, $\text{SNR}$, see Fig.~\ref{fig2}d-f. The measurement signal after an integration time $\tau$ consists of two Gaussian distributions corresponding to the ground and excited qubit states, with mean values $\moy{I_\text{g,e}}$, standard deviations $\moy{\sigma_\text{g,e}}$, and $\text{SNR}^2=(\moy{I_\text{g}}-\moy{I_\text{e}})^2/(\sigma_\text{g}^2+\sigma_\text{e}^2)$, see Fig.~\ref{fig2}d. Experimentally, we prepare the qubit in either its ground or excited state, and measure the rate at which the SNR grows as a function of integration time ($\tau\gg1/\kappa$), yielding the measurement rate $\Gamma_m = \text{SNR}^2/4\tau$ \cite{Didier2015FastInteraction}. In Fig.~\ref{fig2}f we show the measurement rate as a function of measurement amplitude and FPJA gain. As expected, the SNR increases linearly with the measurement amplitude. As the FPJA gain increases, the measurement rate approaches the measurement-induced dephasing rate (dashed line), a hallmark of a highly efficient measurement.

\begin{figure}[h]
	\includegraphics[scale=1.0]{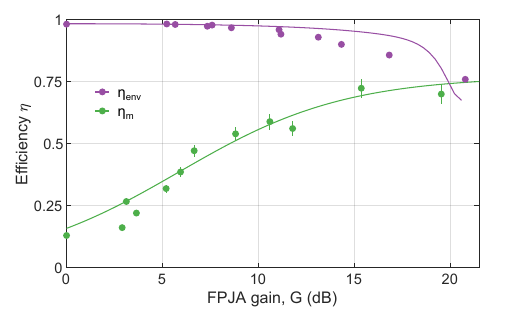}
	\caption{Measurement efficiency $\eta_\text{m}$, in green, and environmental efficiency $\eta_\text{env}$, in purple, as a function of amplifier gain. The solid green line is the prediction of an empirical noise model for the measurement efficiency and the solid purple line is the theoretical prediction for the environmental efficiency (see Supp. Inf.). At high FPJA gain, the measurement efficiency clearly exceeds the $50\%$ theoretical limit for a phase-insensitive amplifier. 
		\label{fig3}}
\end{figure}

The ratio of the measurement rate to the measurement-induced dephasing rate yields the measurement efficiency, $\eta_\text{m} = \Gamma_\text{m}/\Gamma_\phi^\text{m}$ \cite{Didier2015FastInteraction}, shown in Fig.~\ref{fig3} as function of FPJA gain. At low FPJA gain, the measurement efficiency is limited by the system noise temperature of the homodyne measurement setup (Supp. Inf.). As the FPJA gain increases, it overwhelms the noise of the homodyne setup and the efficiency increases. Concurrently the thermal occupancy of the cavity, $n_\text{env}$, increases slightly, as discussed previously in Fig.~\ref{fig2}. This results in an increase of the photon shot-noise in the cavity and therefore of the qubit dephasing, which can be cast as an environmental efficiency \cite{Vijay2012} $\eta_\text{env}=1/(1+2n_\text{env})$, shown in Fig.~\ref{fig3}. At an optimal gain of approximately $15~\text{dB}$, we reach an efficiency of $\eta_\text{m} = 72\pm4\%$ and an environmental efficiency $\eta_\text{env}=88\pm2\%$. An empirical noise model suggests that $\eta_\text{m}$ is primarily limited by residual coupling of the amplifier resonance to the output port \cite{Lecocq2020MicrowaveAmplifier}. Both the measurement inefficiency $1-\eta_\text{m}$ and thermal occupancy of the cavity $n_\text{env}$ can be reduced by at least an order of magnitude with a straightforward redesign of the \red{external coupling to} each resonator of the FPJA \cite{Lecocq2020MicrowaveAmplifier}. \blue{We note that the small discrepancy between the measured backaction and theoretical predictions is potentially due to a weak measurement performed by the amplifier even in the absence of cavity displacement. Indeed, the state-dependant frequency shift of the readout cavity results in a state-dependant gain profile of the amplifier. As a consequence, the qubit state is encoded in the output noise spectrum which, if unmonitored, results in spurious dephasing. This form of backaction has been observed in other experiments and is an active area of research \cite{Eddins2019High-EfficiencyAmplifier,Liu2020NoiseInterferometer}.}

At the optimal FPJA gain, we demonstrate the ability to perform strong projective qubit measurements. We extract a state-estimation fidelity of $97\%$ in $350~\si{\nano\second}$ for $|\alpha|^2\approx2.5$, mostly dominated by the ratio of the qubit relaxation rate and the cavity linewidth (Supp. Inf.). The dynamic range of the amplifier is large enough for a measurement strength $|\alpha|^2>5$, approaching the limits of the dispersive approximation \cite{Blais2004}, inducing spurious qubit transitions.

In conclusion, we have demonstrated a highly efficient measurement of a transmon qubit using a parametrically-driven, directional, phase-sensitive amplifier connected to the readout cavity \red{without an intermediate circulator}. As in other work \cite{Thorbeck2017ReverseAmplifier,Eddins2019High-EfficiencyAmplifier,Rosenthal2020EfficientDetector,Abdo2020On-chipMeasurements}, our approach strives to achieve a high level of integration with superconducting quantum devices. \red{While direct on-chip integration could be feasible, a flip-chip \cite{Rosenberg20173DQubits} or wireless coupling \cite{Gao2019EntanglementInteraction} approach would enable separate optimization of the amplifier and qubit chips.} \red{Straightforward adjustments of} device parameters will enable measurement efficiencies approaching $100\%$, creating a testbed for exploring quantum control protocols based on weak continuous measurement. In addition, we emphasize that the FPJA can be dynamically reconfigured, enabling complex time-domain protocols with tunable control, coupling and measurement of multiple quantum systems.

\red{We thank E. Rosenthal and B. Hauer for helpful discussions and comments on the manuscript.}

%\bibliography{bib_mendeley_fql}% Produces the bibliography via BibTeX.


\end{document}

\title{Supplementary information for "Efficient qubit measurement with a nonreciprocal microwave amplifier"}

\author{F. Lecocq$^{1,2}$, L. Ranzani$^3$, G. A. Peterson$^{1,2}$, K. Cicak$^1$, X. Y. Jin$^{1,2}$, R. W. Simmonds$^1$, J. D. Teufel$^1$ and J. Aumentado$^1$}

\affiliation{$^1$National Institute of Standards and Technology, 325 Broadway, Boulder, CO 80305, USA}
\affiliation{$^2$Dept. of Physics, University of Colorado, 2000 Colorado Ave., Boulder, CO 80309, USA}
\affiliation{$^3$Raytheon BBN Technologies, Cambridge, MA 02138, USA}

\date{\today}% It is always \today, today,

\maketitle

\renewcommand{\thefigure}{S\arabic{figure}}
\setcounter{figure}{0} 

\subsection{Complete system description and amplifier tuning}

\begin{figure*}
	\includegraphics[scale=1.0]{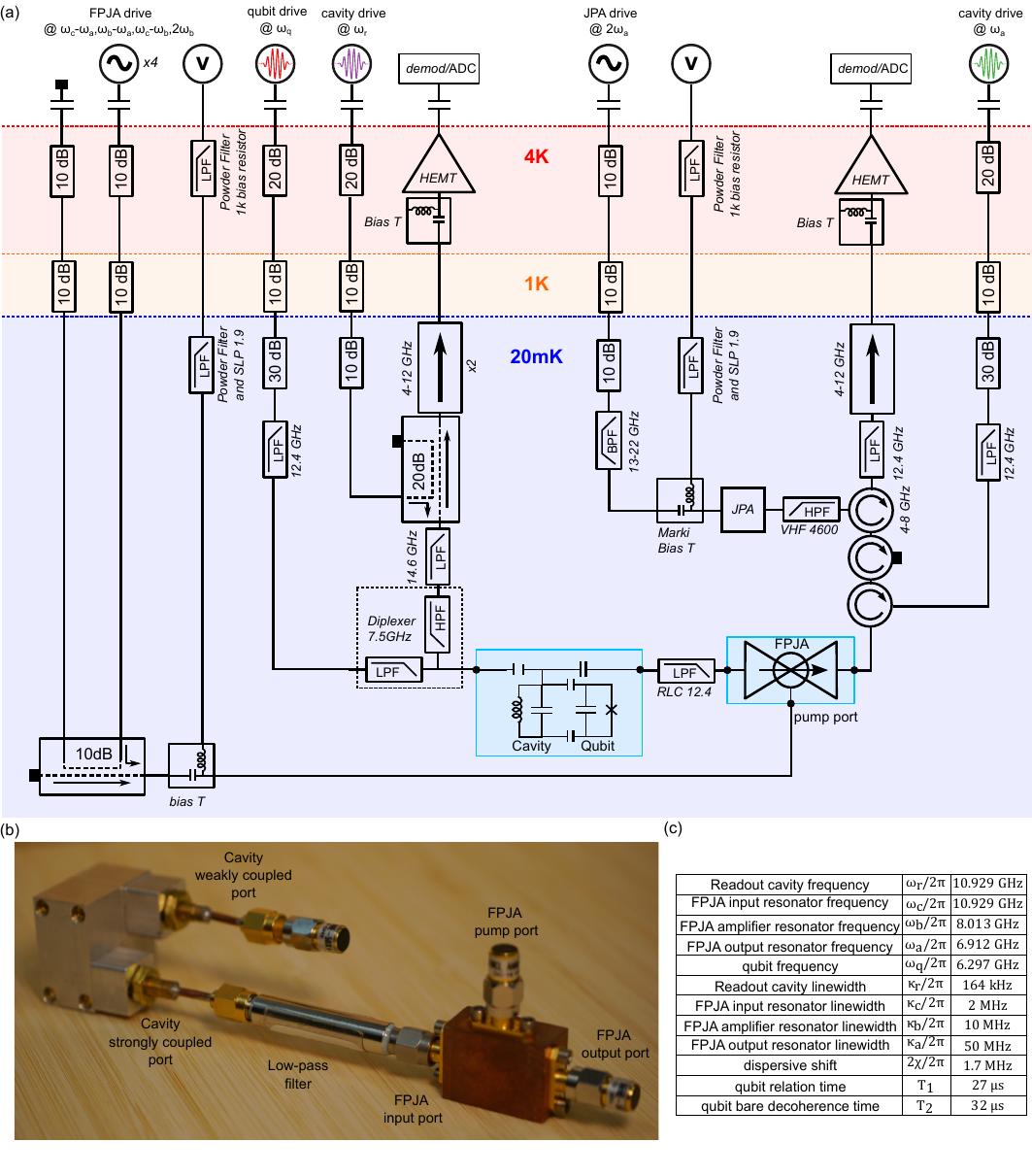}
	\caption{(a) Experimental diagram. (b) Picture of the cavity/FPJA assembly, disconnected from the full experimental setup. (c) table of the main experimental parameters.
	\label{SupFig_DRsetup}}
\end{figure*}

A detailed diagram for the experimental setup and a device picture are shown in Fig.~\ref{SupFig_DRsetup}. The FPJA is the exact same chip used in previous work \cite{Lecocq2020MicrowaveAmplifier}. The readout cavity is addressed via a weakly coupled port and a strongly coupled port. The weakly coupled port is primarily used for driving the qubit and to probe the cavity without using the FPJA. The strongly coupled port is connected to the FPJA, without any circulator in between. \red{In addition to existing on-chip FPJA filtering}, we inserted a low-pass filter with a $12.4~\si{\giga\hertz }$ cutoff to protect the cavity from the amplification drive of the FPJA. In future designs, \red{this additional} low-pass filter could be directly integrated into the FPJA \red{circuit}. A picture of the cavity/FPJA assembly is shown in Fig.~\ref{SupFig_DRsetup}b (for a size reference, all the ports are standard SMA connectors).

\begin{figure*}
	\includegraphics[scale=1.0]{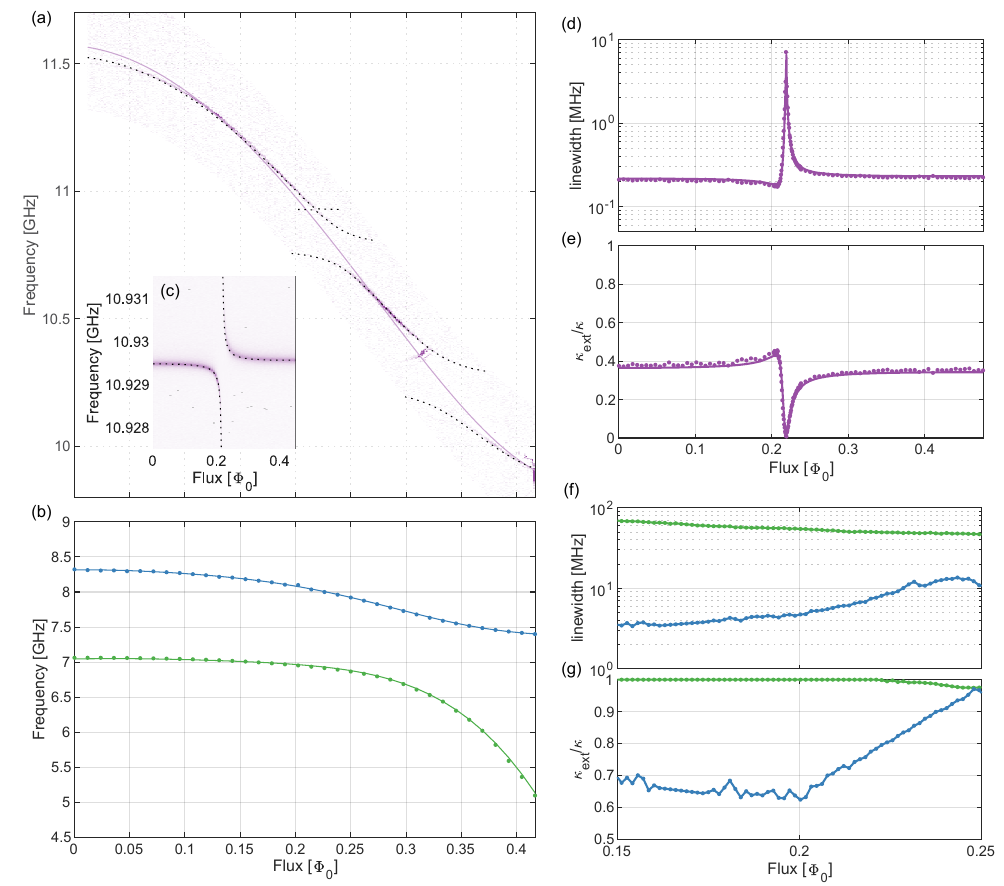}
	\caption{ (a) Magnitude of the reflection measurement off of the output mode, in presence of a weak frequency conversion drive to the input mode (see text), shown as a color scale. The solid purple line is the bare resonance frequency measured when the FPJA input port is connected to a $50~\si{\ohm}$ load. The dashed black line is the prediction from a full model including the readout cavity and spurious resonances due to the length of transmission line between the readout cavity and the FPJA. (b) Resonance frequency of the output resonator $a$ (green) and amplification resonator $b$ (blue). (c) Magnitude of the reflection measurement off of the readout cavity shown as a color scale, showing an anti-crossing when the input of the FPJA and the readout cavity are in resonance. The dashed lines are the predictions from the same model than in (a). The FPJA input mode in resonance with the readout cavity for $\Phi\approx0.219\Phi_0$. (d) Linewidth, $\kappa$, of the readout cavity and (e) ratio of the external coupling rate, $\kappa_\text{ext}$, to the total linewidth of the cavity. Solid lines are predictions from the model. (f) Linewidth of the output resonator (green) and amplification resonator (blue) and (g) ratio of the external coupling rate to the total linewidth for these resonators. 
	\label{SupFig_ModCurve}} 
\end{figure*}

In Fig.~\ref{SupFig_ModCurve} we show the resonance frequencies and linewidths of the FPJA and the readout cavity, as a function of the flux bias, $\Phi$, through the SQUID of the FPJA. The frequencies and linewidth of the output resonator $a$ and amplification resonator $b$ are directly accessed from a reflection measurement off the FPJA output port, shown in Fig.~\ref{SupFig_ModCurve}b, f and g. The input mode $c$ is not directly accessible due to the on-chip FPJA filters. However, reflection measurements off of the output mode $a$ in the presence of a single weak frequency conversion drive at the frequency difference $\omega_c(\Phi)-\omega_a(\Phi)$ allows us to recover the frequency dependence of the input mode $c$, shown in Fig.~\ref{SupFig_ModCurve}a. The data (purple color scale) deviates from the bare resonance frequency measured when connected to a $50~\si{\ohm}$ (purple line) load \cite{Lecocq2020MicrowaveAmplifier}, and this effect is due to the presence of standing waves between the cavity and the FPJA. Multiple anti-crossings are visible, and the data is well explained by including spurious resonances with a free spectral range of $535~\si{\mega\hertz}$ and adjusting the coupling to each of these resonances (dashed black lines). The readout cavity is coupled to the same spurious resonances, leading to an avoiding crossing with the input resonator $c$ around $10.93~\si{\giga\hertz}$ and $\Phi\approx0.219\Phi_0$. The same anti-crossing is revealed by a reflection measurement off of the weakly coupled port of the readout cavity, shown in Fig.~\ref{SupFig_ModCurve}c, and the measured linewidths are shown in Fig.~\ref{SupFig_ModCurve}d and e. The complete model agrees well with both the frequency dependence and linewidths (solid lines in Fig.~\ref{SupFig_ModCurve}d and e).

We tune the amplifier by comparing measured scattering parameters to theoretical predictions, like in previous work \cite{Lecocq2020MicrowaveAmplifier}, and using the independently measured resonance frequencies and linewidths. We first calibrate each parametric drive power separately (three frequency conversion drives and one parametric amplification), using the SQUID flux as the single free parameters to account for frequency rectification.

We emphasize here that the parametric pumps couple the readout cavity to the external ports of the FPJA. For all the pump configurations discussed in this work, the cavity response is well approximated by a Lorentzian and its effective linewidth $\kappa$ can be extracted by measuring the reflection coefficient from the weakly coupled port of the cavity. In particular, when the FPJA is programmed as a directional phase sensitive amplifier, the effective linewidth of the readout cavity is dominated by the frequency conversion processes toward the output port, yielding a highly overcoupled resonance with  $\kappa\approx2\pi\times2.58~\si{\mega\hertz}\gg\kappa_\text{r}$.

\red{Finally, we note that the qubit readout could also be performed in transmission (from the weakly coupled port of the cavity to the FPJA ouput port) but with the added technical complexity of frequency conversion from $\omega_c$ to $\omega_a$ which requires different drive and local oscillator frequencies.}

\subsection{Qubit coherence}
In this section we discuss the qubit coherence time as a function the mode of operation of the FPJA. For all modes of operation, we measure the same qubit relaxation time $T_1 = 27~\si{\micro\second}$.

When the FPJA is largely detuned from the readout cavity ($\Phi=0$) and unpumped, it can be seen as an open circuit. We use the weakly coupled cavity port to extract a baseline for the qubit coherence time, $T_2^\text{off}  = 32~\si{\micro\second}$. Using Eq.~1 in the main text and the cavity linewidth measured in \ref{SupFig_ModCurve}, we extract a thermal cavity occupancy of $n_\text{env}=0.01$.

We then tune the FPJA into resonance with the cavity and turn on the frequency conversion pump between the input and output mode, as described in the previous section. We then measure the qubit coherence time, $T_2^\text{FC}  = 17~\si{\micro\second}$. Interestingly, due to an increase of the effective cavity linewidth, we extract an identical a thermal cavity occupancy of $n_\text{env}=0.01$.

\begin{figure}
	\includegraphics[scale=1.0]{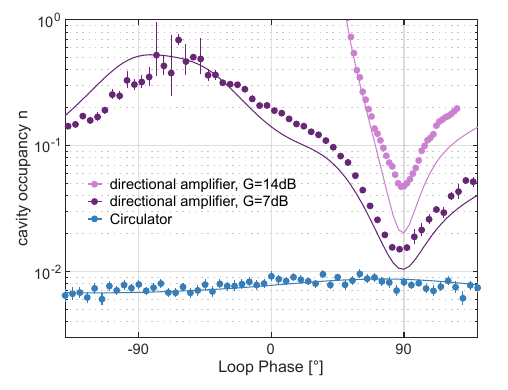}
	\caption{Cavity occupancy as a function of loop phase for the FPJA programmed as a circulator (blue) and a directional phase-sensitive amplifier with two different gains (dark and light purple). Solid lines are theoretical predictions.
	\label{SupFig_NvsPhase}}
\end{figure}

We then program the FPJA as a circulator and the measured qubit coherence time is shown in Fig.~\ref{SupFig_NvsPhase} as a function of loop phase \cite{Lecocq2017NonreciprocalAmplifier}, in good agreement with theoretical predictions where each bath temperature is set to $n_\text{env}=0.01$. Similarly, good agreement is observed when the FPJA is programmed as a directional phase-sensitive amplifier.

%subsubsection{Amplifier compression}

%In Fig.~\ref{SupFig_Compression} we show the measured amplifier compression point expressed in terms of number of photon in the readout cavity.

%\begin{figure}
%	\includegraphics[scale=1.0]{./SupFig_Compression.pdf}
%	\caption{\textbf{Amplifier compression}. The dashed line is the JPA compression divided by the FPJA gain. They compress together.
%	\label{SupFig_Compression}}
%\end{figure}

\subsection{System noise model}

\begin{figure}
	\includegraphics[scale=1.0]{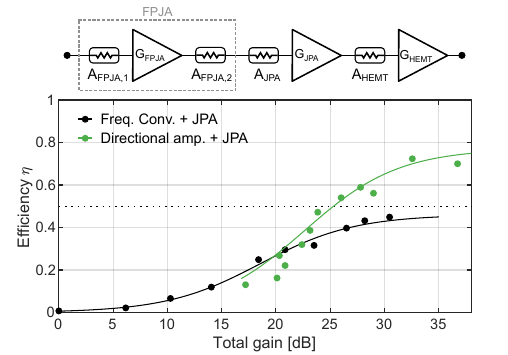}
	\caption{Measured efficiency as a function of the total quadrature gain (sum of FPJA gain and JPA gain). The FPJA is either programmed as a frequency converter with a variable JPA gain (black) or programmed as a variable gain directional phase sensitive amplifier with a fixed JPA gain of $18.2~\text{dB}$ (green, same data as in Fig.~\ref{fig3}). The solid line are predictions of an empirical noise model (see text).
		\label{SupFig_NoiseModel}}
\end{figure}

In this section we describe the empirical model used for predicting the measurement efficiency in Fig.~\ref{fig3}, based on a chain of cascaded amplifiers, as shown in Fig.~\ref{SupFig_NoiseModel}. We estimate the system noise of the chain, $n_\text{sys}$, and convert it into an efficiency $\eta_\text{m}=1/(1+2n_\text{sys})$.

First we set the FPJA as a simple frequency converter between the input and output ports ($G_\text{FPJA}=1$). At high JPA gain, the total system noise is limited by the power loss before the JPA, $A_\text{FPJA,1}A_\text{FPJA,2}A_\text{JPA}$. Reflection measurements off of the FPJA show a power loss of about $0.68$ or $-1.7~\text{dB}$, out of which half occurs each way, allowing us to extract $A_\text{FPJA,1}A_\text{FPJA,2}=0.82$. From the maximum efficiency measured at high gain, we can then extract the loss between the FPJA and the JPA $A_\text{JPA}=0.56$. Finally, from the efficiency at zero JPA gain we extract the added noise of the chain after the JPA, $n_\text{sys}^\text{HEMT}=36$. This model is shown as the black solid line in Fig.~\ref{SupFig_NoiseModel}.

We now program the FPJA as a directional phase-sensitive amplifier. In absence of FPJA gain, we independently measure a reduction of the transmission from the weakly coupled of the cavity to the output port of the FPJA by $0.72$ or $-1.4~\text{dB}$. This is due to residual loss in the amplification mode $b$. In this case we now have $A_\text{FPJA,1}A_\text{FPJA,2}=0.59$. Assuming the loss is spread equally before and after the gain $A_\text{FPJA,1}=A_\text{FPJA,2}$ we obtain the solid green line in Fig.~\ref{SupFig_NoiseModel}, in good agreement with the data.

\subsection{Measurement fidelity}
\begin{figure}
	\includegraphics[scale=1.0]{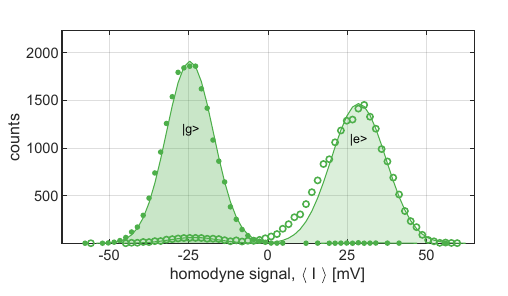}
	\caption{Qubit measurement fidelity. Histogram of $2\times10^4$ measurements integrated over $350~\si{\nano\second}$ with $|\alpha|^2\approx2.5$, for the qubit prepared in its ground state (circles) or excited state (crosses). The JPA gain is set at $18~\text{dB}$ and the FPJA gain  at $12~\text{dB}$.  We extract a measurement fidelity of $97\%$.
	\label{SupFig_Fidelity}}
\end{figure}

In Fig.~\ref{SupFig_Fidelity} we show the histogram of $2\times10^4$ measurements integrated over $350~\si{\nano\second}$ with $|\alpha|^2\approx2.5$, for the qubit prepared in its ground state (dots) or excited state (circles). The JPA gain is set to $18~\text{dB}$ and the FPJA gain to $12~\text{dB}$.  We extract a measurement fidelity of $\mathcal{F}=1-P(e|g)-P(g|e)\approx97\%$, where $P(x|y)$ is the probability of measuring the qubit in the state $x$ when prepared in the state $y$. \red{We estimate the following contributions to the measurement infidelity ($1-\mathcal{F}\approx3\%$): overlap between the distributions ($0.1\%$), qubit thermal population ($0.3\%$), relaxation during the measurement ($1.5\%$ to $2\%$) and state preparation error ($0.6\%$ to $1.1\%$).}

